\documentclass[runningheads]{svmult}

\usepackage{makeidx}   
\usepackage{graphicx}  
\usepackage{subeqnar}  
\usepackage{multicol}  
\usepackage{physprbb}  
\makeindex             

\newcommand{\greeksym}[1]{{\usefont{U}{psy}{m}{n}#1}}
\newcommand{\umu}{\mbox{\greeksym{m}}}

\begin{document}
%
\title*{Spheroidal Galaxies/QSOs Connection}

%
%
%
\toctitle{Spheroidal Galaxies/QSOs Connection}
%
%
\titlerunning{Spheroidal Galaxies/QSOs Connection}
%
\author{Luigi Danese\inst{1}
\and Gian Luigi Granato\inst{2} \and Laura Silva\inst{3} \and
Manuela Magliocchetti\inst{1} \and Gianfranco De Zotti\inst{2}} \
\authorrunning{L. Danese et al.}
%
%
\institute{SISSA/ISAS, Via Beirut 2--4, I-34014 Trieste, Italy
\and Osservatorio Astronomico di Padova, Vicolo dell'Osservatorio
5, I-35122 Padova, Italy \and Osservatorio Astronomico di Trieste,
Via Tiepolo 11, I-34131 Trieste, Italy }

\maketitle              

\def\lsim{\, \lower2truept\hbox{${< \atop\hbox{\raise4truept\hbox{$\sim$}}}$}\,}
\def\gsim{\, \lower2truept\hbox{${> \atop\hbox{\raise4truept\hbox{$\sim$}}}$}\,}

\begin{abstract}
In view of the extensive evidence of a tight inter-relationship
between spher\-oidal galaxies (and galactic bulges) and massive
black holes hosted at their centers, a consistent model must deal
jointly with the evolution of the two components. We describe one
viable model, which successfully accounts for the local luminosity
function of spheroidal galaxies, their photometric and chemical
properties, deep galaxy counts in different wavebands, including
those in the (sub)-mm region which proved to be critical for
current semi-analytic models stemming from the standard
hierarchical clustering picture, clustering properties of SCUBA
galaxies, of EROs, and of LBGs, as well as for the local mass
function of massive black holes and for quasar evolution.
Predictions that can be tested by surveys carried out by SIRTF are
presented.
\end{abstract}

\section{Introduction}
The hierarchical clustering model with a scale invariant spectrum
of density perturbations in a Cold Dark Matter (CDM) dominated
universe has proven to be remarkably successful in matching the
observed large-scale structure as well as a broad variety of
properties of galaxies of different morphological types (e.g.
\cite{Cole2000,Granato2000}). However, serious shortcomings of
this scenario have also become evident in recent years. The
critical point can be traced back to the relatively large amount
of power on small scales predicted by this model which would imply
far more dwarf galaxies or substructure clumps within galactic and
cluster mass halos than are observed (the so-called ``small-scale
crisis''
(\cite{Haiman2001,Somerville2001,KamionkowskiLiddle2000,Mooreetal1999}),
unless star formation in small objects is strongly suppressed (or
the small scale power is reduced by modifying the standard model).

At the other extreme of the galaxy mass function we have another
strong discrepancy with model predictions, that we might call
``the massive galaxy crisis'': even the best semi-analytic models
(\cite{Granato2000,Devriendt2000}) hinging upon the standard
picture for structure formation in the framework of the
hierarchical clustering paradigm, fall short by a substantial
factor (up to about 10) to account for the (sub)-mm (SCUBA and
MAMBO) counts of galaxies, most of which are probably massive
objects undergoing a very intense star-burst (with star formation
rates $\sim 1000\,\hbox{M}_\odot\,\hbox{yr}^{-1}$) at $z>2$ (see,
e.g. \cite{Dunlop2001}). Recent optical data confirm that most
massive ellipticals were already in place and (almost) passively
evolving up to $z\simeq 1$--1.5, implying that they were fully
assembled by $z \sim 2.5$, although the issue is still somewhat
controversial
(\cite{RenziniCimatti1999,Ferguson2000,Daddi2000b,Martini2001,Rodighiero2001,Cohen2001,Im2001,McCarthy2001}).
These data are more consistent with the traditional ``monolithic''
approach whereby giant ellipticals formed most of their stars in a
single gigantic starburst at substantial redshifts, and underwent
essentially passive evolution thereafter.

On the contrary, in the canonical hierarchical clustering paradigm
the smallest objects collapse first and most star formation
occurs, at relatively low rates, within relatively small
proto-galaxies, that later merged to form larger galaxies. Thus,
the expected number of galaxies with very intense star formation
is far less than detected in SCUBA and MAMBO surveys and the
surface density of massive evolved ellipticals at $z\gsim 1$ is
also smaller than observed. The ``monolithic'' approach, however,
is inadequate to the extent that it cannot be fitted in a
consistent scenario for structure formation from primordial
density fluctuations.

\section{Relationships between quasar and galaxy evolution}
The above difficulties, affecting even the best current recipes,
may indicate that new ingredients need to be taken into account. A
key new ingredient may be the mutual feedback between formation
and evolution of spheroidal galaxies and of active nuclei residing
at their centers
(\cite{KormendyRichstone1995,Magorrian1998,HallGreen1998,vanderMarel1999,FerrareseMerritt2000,Gebhardt2000a,Gebhardt2000b,McLureDunlop2001,MerrittFerrarese2001a,MerrittFerrarese2001b}).
In this framework, \cite{Granato2001} elaborated the following
scheme (see also \cite{EalesEdmunds1996,SilkRees1998,Monaco2000}):

\begin{itemize}

\item Feed-back effects, from supernova explosions and from active
nuclei (note that supernova feedback alone falls short of solving
the dearth of dwarf galaxies, \cite{MacLowFerrara1999}, but
photo-ionization by the UV background re-ionizing the
inter-galactic medium (IGM) could do the job
\cite{Somerville2001}) delay the collapse of baryons in smaller
clumps while large ellipticals form their stars as soon as their
potential wells are in place; {\it the canonical hierarchical CDM
scheme -- small clumps collapse first -- is therefore reversed for
baryons}.

\item Large spheroidal galaxies therefore undergo a phase of high (sub)-mm
luminosity.

\item At the same time, the central black-hole (BH) grows by accretion and the quasar
luminosity increases; when it reaches a high enough value, its
action (ionization and heating of the gas) stops the star
formation and eventually expels the residual gas. This explains
the observed correlation between BH and host spheroidal masses
(see \cite{SilkRees1998,Fabian1999}). The same mechanism
distributes in the IGM a substantial fraction of metals and may
pre-heat the IGM. The onset of quasar activity (and the
corresponding squelching of star formation) occurs earlier for
more massive objects. The duration of the star-burst increases
with decreasing mass from $\sim 0.5$ to $\sim 2\,$Gyr.

\item This implies that the star-formation
activity of the most massive galaxies quickly declines for $z\lsim
3$, i.e. the redshift distribution of SCUBA/MAMBO galaxies should
peak at $z\gsim 3$, just before quasars reach their maximum
luminosity (at $z\simeq 2.5$). This:

\begin{itemize}

\item explains why very luminous quasars are more easily detected at (sub)-mm
wavelengths for $z \gsim 2.5$ \cite{Omont2001}. The latter authors
argue that a large fraction of the observed (sub)-mm emission is
powered by a starburst, as expected from this model

\item implies an extremely steep (essentially exponential) decline
of (sub)-mm counts at bright fluxes, as indicated by recent data.

\end{itemize}

\item A ``quasar phase'' follows, lasting $10^7$--$10^8\,$yrs.

\item A long phase of passive evolution of galaxies ensues, with their colors becoming
rapidly very red [Extremely Red Object (ERO) phase].

\item Intermediate- and low-mass spheroids have lower Star Formation Rates and less
extreme optical depths. They show up as Lyman-Break Galaxies
(LBGs).

\item Therefore, in this scenario, large ellipticals evolve essentially as in the
''monolithic'' scenario, yet in the framework of the standard
hierarchical clustering picture.

\end{itemize}

\noindent The various aspects and implications of this compound
scheme have been addressed by our group in a series of papers.
\cite{Salucci1999} estimated the mass function of quiescent BHs at
the centers of local galaxies. They found a dichotomy between
large and small BHs: the former are hosted in elliptical galaxies
and tend to be not obscured, while the latter, found in the bulges
of spiral galaxies, can be reactivated by interactions and are
frequently obscured. Their results are consistent with the high
luminosity quasar activity occurring as a single short-lived event
at relatively high redshift with luminosity close to the Eddington
limit. Later, lower luminosity, nuclear activity may be due to
re-activation of the central BH, due, e.g., to interactions.
\cite{Monaco2000} analyzed the evolution of the quasar luminosity
function in the framework of a model in which the spheroidal
galaxies (and bulges of later-type galaxies) and the BHs at their
centers form and evolve in parallel. Adopting the standard
\cite{PressSchechter1974} formation rate of dark matter halos,
they found that consistency with the data is achieved if, and only
if, BHs in bigger galactic halos form earlier. In other words, the
interval between the onset of star formation and the peak of
quasar luminosity is shorter for the more massive objects. Some of
the other, most recent, results are briefly described in the next
section.

\begin{figure}[t]
\begin{center}
\vspace{-5mm}
\includegraphics[width=.9\textwidth,height=6cm]{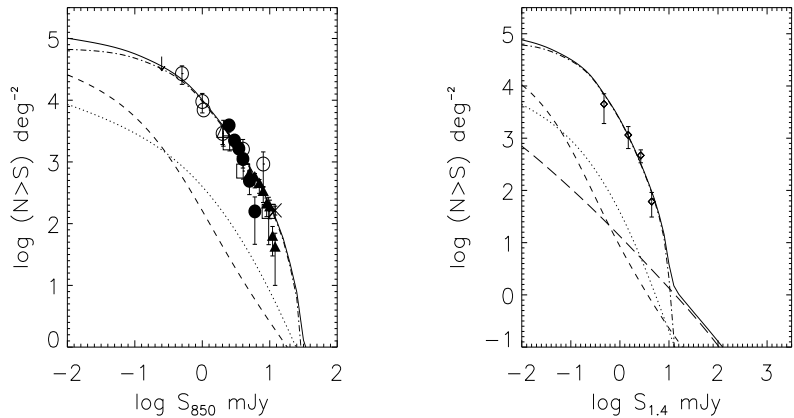}
\end{center}
\vspace{-5mm} \caption[]{Integral source counts at $850\,\mu$m
(left panel) and at $1.4\,$mm (right panel) predicted by the model
by \cite{Granato2001} compared with observations. The dotted,
dashed and dot-dashed lines show the contributions of starburst,
spiral, and forming elliptical galaxies, respectively. The
long-dashed line (shown only at $1.4\,$mm) gives the counts of
radio sources, after \cite{toffolatti}. The solid line shows the
total counts. References for the data points can be found in
\cite{Perrotta2001b}} \label{counts850}
\end{figure}

\section{Some recent results and predictions}

\subsection{Counts at (sub)-mm wavelengths}
The (sub)-mm counts are expected to be very steep because of the
combined effect of the strong cosmological evolution of dust
emission in spheroidal galaxies and of the strongly negative
K-correction (the dust emission spectrum steeply rises with
increasing frequency). The model by \cite{Granato2001} has extreme
properties in this respect: above several mJy its $850\,\mu$m
counts reflect the high-mass exponential decline of the mass
function of dark halos. In this model, SCUBA/MAMBO galaxies
correspond to the phase when massive spheroids formed most of
their stars at $z\gsim 2.5$; such objects essentially disappear at
lower redshifts. On the contrary, the counts predicted by
alternative models (which are essentially phenomenological
\cite{Blain1999,Tan1999,RowanRobinson2001,Pearson2001,Takeuchi2001}),
while steep, still have a power law shape, and the redshift
distribution has an extensive low-$z$ tail.

As illustrated by Fig.~1, the recent relatively large area surveys
\cite{Scott2001,Borys2001} are indeed suggestive of an exponential
decline of the $850\,\mu$m counts above several mJy. Further
evidence in this direction comes from MAMBO surveys at $1.2\,$mm
\cite{Bertoldi2000}; see \cite{Perrotta2001b}).

\subsection{Lensing effects on (sub)-mm counts}
A direct consequence of the extreme steepness of the (sub)-mm
counts predicted by this model is that their bright tail is
strongly affected by gravitational lensing
\cite{Perrotta2001a,Perrotta2001b}. In fact, although the
probability of strong lensing is very small, it has a power-law
tail ($p(A) \propto A^{-3}$) extending up to large values of the
magnification $A$. Thus, if counts are steep enough, the fraction
of lensed sources at bright fluxes may be large.

\cite{Perrotta2001a} and \cite{Perrotta2001b} find that, according
to the model by \cite{Granato2001}, essentially all
proto-spheroidal galaxies brighter than $S_{850\mu{\rm m}} \simeq
60$--70$\,$mJy are gravitationally lensed. Allowing for the other
populations of sources contributing to the bright mm/sub-mm
counts, they find that the fraction of gravitationally lensed
sources may be $\simeq 40\%$ at fluxes slightly below
$S_{850\mu{\rm m}} = 100\,$mJy. If so, large area surveys such as
those to be carried out by {\sc Planck}/HFI or by forthcoming
balloon experiments like BLAST and ELISA will probe the large
scale distribution of the peaks of the primordial density field.
For comparison, the maximum fraction of lensed sources predicted
by current phenomenological models is $\lsim 5\%$.

\subsection{Clustering of SCUBA/MAMBO galaxies}
Since, in the above scheme, SCUBA galaxies correspond to the rare,
massive density peaks at high-$z$, they are expected to be highly
biased tracers of the dark matter distribution, and therefore to
be strongly clustered. The recent analyses by
\cite{Magliocchetti2001} and \cite{Perrotta2001b} have shown that
the implied angular correlation function is indeed consistent with
the results by \cite{Giavalisco1998} on the clustering of LBGs,
and by \cite{Scott2001} and \cite{Peacock2000} on clustering of
SCUBA galaxies. The estimated $w(\theta)$ for bright SCUBA
galaxies \cite{Scott2001} and for EROs \cite{Daddi2000a} indicate
dark halo masses $~ 10^{13}\,\hbox{M}_\odot$ for these sources;
the $w(\theta)$ for LBGs \cite{Giavalisco1998} indicate typical
masses at least 10 times lower (see also \cite{Moustakas2001}).

\subsection{Chemical and photometric evolution}
The problem has been addressed by \cite{Romano2001}. The short
duration of the star-burst in spheroidal galaxies, and its
occurrence at $z\gsim 3$, assumed by the model, are consistent
with the supra-solar $Mg/Fe$ ratio in ellipticals, with the
tightness of their fundamental plane and of their color--$\sigma$
relation. But the model also compares successfully with a broad
variety of detailed observational data: the correlation between
the abundances of $Fe$ and those of ``enhanced'' elements ($C$,
$N$, $O$, $Ne$, $Mg$, $Si$, $S$), the correlation between $Fe$ and
$Mg_2$, the template Spectral Energy Distribution of local
ellipticals, the color ($V-K$)-magnitude ($M_V$) and the
$(J-K)$--$(V-K)$ relations. The key ingredients allowing to match
the observational data, and in particular the relationships among
abundances of the various elements are, on one side, the decrease
of the duration of the star-burst with increasing galactic mass,
and, on the other side, the fact that the effect of stellar
feedback is higher in less massive objects, so that the amount of
material enriched by SNII that can be retained is lower. The main
difference with the majority of previous chemical evolution models
is that star formation is stopped not by supernova-driven winds
(which also fall short of providing enough preheating of the
proto-cluster medium,
\cite{ValageasSilk1999,Balogh2001,KravtsovYepes2000}), but by the
energy injected by the active nucleus.

Note that this model also accounts for the roughly solar, or
supra-solar, metallicity of the quasar broad-line regions and of
supra-solar $N/C$ and $Fe/\alpha$ ratios in the more luminous
objects observed out to $z >4$ \cite{HamannFerland1999} and even
up to $z\sim 6$ \cite{Pentericci2001}. This implies that the
medium must have been chemically enriched before the quasars
became observable and on short timescales ($\le 1\,$Gyr, at least
at the highest redshifts).

Another interesting implication of the model, also discussed by
\cite{Romano2001}, is the dependence of the ratio $M_{\rm dark
matter}/M_{\rm stars}$ on $M_{\rm stars}$. This ratio has a
minimum value $\simeq 20$ for $M_{\rm stars}\simeq
10^{10}$--$10^{11}\,\hbox{M}_\odot$, in good agreement with the
results by \cite{McKay2001}, and increases at smaller masses (in
keeping with the findings by \cite{Persic1996}) due to the
increasing fraction of gas expelled by supernova feedback, as well
as at higher masses, due to the increase of the cooling time of
the gas in the outer regions of the galactic halo.

\begin{figure}[t]
\begin{center}
\vspace{-5mm}
\includegraphics[width=.9\textwidth,height=6cm]{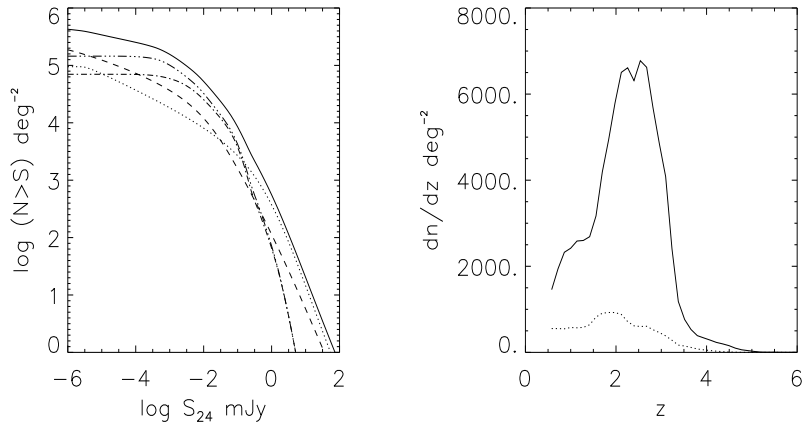}
\end{center}
\vspace{-5mm} \caption[]{Integral source counts (left-hand panel)
and redshift distribution (right-hand panel) for a flux limit of
$30\mu$Jy (right panel) at $24\,\mu$m predicted by the model by
\cite{Granato2001}. In the left-hand panel, the dotted, dashed and
dot-dashed lines show the contributions of starburst, spiral, and
forming elliptical galaxies, respectively, while the
three-dots/dashed line shows the total counts of ellipticals,
including also those where the star-formation has ended. In the
right-hand panel, the solid and the dotted lines show the redshift
distributions of ellipticals during and after the star-formation
phase, respectively} 
\label{counts24micron}
\end{figure}

\subsection{Predictions for SIRTF surveys}
SIRTF surveys have the potential of providing further tests of the
model. In particular, the GOODS
(http://www.stsci.edu/science/goods) $24\,\mu$m survey should
reach the confusion limit at 30--$100\,\mu$Jy. According to the
model, about 50\% of the detected galaxies should be spheroidal
galaxies forming their stars at $z \gsim 2$. About 400--600 such
objects are expected over an area of 0.1 square degree (see
Fig.~2). Their redshift distribution is predicted to peak at $z$
slightly above 2, with a significant tail extending up to $z\gsim
3$.

\bigskip\noindent
{\it Acknowledgements.} We benefited from many helpful exchanges
with C. Baccigalupi, F. Matteucci, F. Perrotta, D. Romano. Work
supported in part by ASI and MIUR.

\end{document}